\documentclass[runningheads]{svmult}

\usepackage{makeidx}   
\usepackage{graphicx}  
\usepackage{subeqnar}  
\usepackage{multicol}  
\usepackage{physprbb}  
\makeindex             

\begin{document}
\title*{Mapping the Stellar Dynamics of M31}
\toctitle{Mapping the Stellar Dynamics of M31}
\titlerunning{Mapping the Stellar Dynamics of M31}
\author{Helen Merrett\inst{1}
  \and M.~Merrifield\inst{1}
  \and K.~Kuijken\inst{2,3}
  \and A.~Romanowsky\inst{1}
  \and N.~Douglas\inst{3}
  \and N.~Napolitano\inst{3}
  \and M.~Arnaboldi\inst{4}
  \and M.~Capaccioli\inst{5}
  \and K.~Freeman\inst{6}
  \and O.~Gerhard\inst{7}
  \and D.~Carter\inst{8}
  \and N.~W.~Evans\inst{9}
  \and M.~Wilkinson\inst{9}
  \and C.~Halliday\inst{10}
  \and T.~Bridges\inst{11}}
\authorrunning{Merrett et al.}

\institute{School of Physics \& Astronomy, University of Nottingham,
  University Park, Nottingham, NG7 2RD, UK
\and Leiden Observatory, Leiden, The Netherlands
\and Kapteyn Institute, Groningen, The Netherlands
\and Osservatorio Astronomico di Torino, Pino Torinese,  Italy
\and Osservatorio di Capodimonte, Naples, Italy
\and Research School of Astronomy and Astrophysics, Australian
  National University, Canberra, Australia
\and Astronomisches Institut, Universit\"{a}t Basel, Switzerland
\and Astrophysics Research Institute, Liverpool John Moores
  University, UK
\and Institute of Astronomy, Cambridge, UK
\and Osservatorio Astronomico di Padova, Padova, Italy
\and Physics Department, Queen's University, Kingston, Canada
}

\maketitle              

\begin{abstract}
  Using the Planetary Nebula Spectrograph, we have observed and
  measured the velocities for 2764 PNe in the disk and halo of
  the Andromeda galaxy.
  Preliminary analysis using a basic ring model shows a rotation
  curve in good agreement with that obtained from {\sc Hi} data out to
  $\sim$\,20\,kpc.
  Some substructure has also been detected within the velocity
  field, which can be modeled as the continuation of the
  tidal--remnant known as the Southern Stream, as it passes through
  Andromeda's disk. 
\end{abstract}

\section{Introduction}
The Andromeda galaxy, the nearest large spiral galaxy ($D\sim
770\,$kpc), provides a unique observational target. M31's size and proximity
allow surface structures to be resolved and studied in unprecedented
detail.
However, due to its large angular size, until now it has not been
practical to perform a large survey of M31's stellar disk dynamics.
Previous dynamical studies have either measured gas in the disk
(e.g. {\sc Hi} \cite{braun91} and {\sc Hii} \cite{rubin70}) or tracer
populations in the halo (e.g. globular clusters \cite{perrett02}; and
RGB stars \cite{reitzel02} and \cite{ibata04}).

With the commissioning of the Planetary Nebula Spectrograph (PN.S) at the
William Herschel Telescope in La Palma, it has become possible to perform
such a survey. Nine nights of observations (13 allocated) with the PN.S and 6
nights surveying the halo with the Wide Field Camera on the Isaac Newton
Telescope have led to a preliminary catalogue of 2764 PNe, delving
$\sim$\,4.5 magnitudes into the PN luminosity function, extending
$\sim$\,2.5$^\circ$ along the major axis, $\sim$\,1.5$^\circ$ along
the minor axis, and covering various elements of substructure in the halo's 
stellar population density, such as the Southern Stream and Northern
Spur \cite{ferg02}.

\section{Rotation}

A raw rotation curve for the PN.S data has been extracted by averaging
PN velocities using a flat ring model (10 concentric rings of width
3.36\,kpc), where PNe are assumed to lie in a thin disk and move on
circular orbits. The PNe within a ring are weighted according to their
angular distance from the major axis ($\phi$) and those that are
closer to the minor axis than the major axis are omitted, as are PNe
close to identified external objects. A
3$\sigma$--clipping routine has been used to eliminate PNe with
discrepant velocities, such as halo, foreground or background
contaminants.  

\begin{figure}[t!]
\begin{center}
\includegraphics[width=0.64\textwidth]{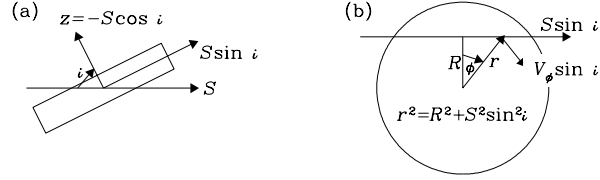}
\end{center}
\caption[]{Model for line--of--sight integration through the disk, shown
  with the disk \textbf{(a)} edge on, \textbf{(b)} face on} 
\label{diag}
\end{figure}

As the Andromeda galaxy is highly inclined ($i=77^\circ$) a correction
to the raw rotation velocities must be made to account for the
line--of--sight integration through the disk. This tends to reduce the
observed velocities, $V_{\phi_\mathrm{obs}}$, with respect to the rotation
velocity, $V_\phi$, by a factor $f(R', z', i)$, where the parameters
are as shown in Fig.\ref{diag} , and
\begin{equation}
   V_{\phi} = \frac{V_{\phi_\mathrm{obs}}}
   {f\left(R',z',i\right)}  
\end{equation}
where
\begin{equation}
  f\left(R',z',i\right) = \frac{\int_0^\infty
  \nu(r',z') \cos\phi \D S'} {\int_0^\infty \nu(r',z') \D S'}.
\end{equation} \label{eqf}
Here, $\nu$ is the number density of PNe, and the primed quantities
have been defined in units of the photometric 
disk scale length, $r_0$, i.e. $R'=\frac{R}{r_0}$, $r'=\frac{r}{r_0}$,
$z'=\frac{z}{r_0}$ and $S'=\frac{S}{r_0}$. 
An expression for $f(R',z',i)$ can be found by combining (\ref{eqf}),
the geometrical relations in Fig.\ref{diag} and the expression 
\begin{equation}
  \nu(r',z') = \nu_0\exp\left(-r'-\frac{z' r_0}{z_0}\right)
\end{equation}
which assumes that the disk is exponentially declining both radially
and perpendicularly to the plane, and $z_0$ is the disk scale height.

A somewhat more significant correction is that of the asymmetric
drift, which accounts for the fact that not all PNe are moving on
circular orbits. The difference between the true circular rotation
velocity, $V_c$, and $V_\phi$, or asymmetric drift, is
related to the velocity dispersion. A correction for this has been
performed using the method in \cite{neistein99}, whereby it is assumed that
\begin{equation}  
  V_c^2 = V_\phi^2 + \sigma_\phi^2 \left(2R'-1\right)
\end{equation}

In order to perform this correction the velocity dispersion has been
fitted as an exponential function of the form
$\sigma_\phi=\sigma_0\E^{-r/2r_0}$. The dynamical scale length has been set to
twice the photometric value to maintain a constant disk scale height
with radius \cite{bottema93}.

\begin{figure}[t!]
\begin{center}
\includegraphics[width=0.80\textwidth]{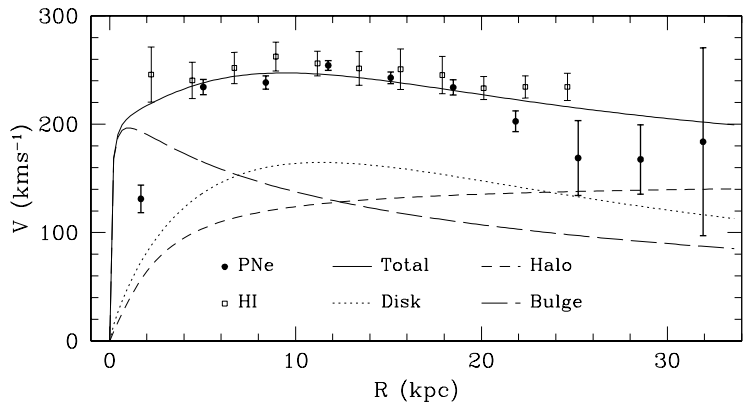}
\end{center}
\caption[]{PNe rotation curve. 
  The filled circles are the PN.S data (binned into 10 concentric
  rings with the PNe nearest the minor axis eliminated), corrected for
  the line--of--sight integration through a disk of scale height $z_0$
  and the asymmetric drift.
  The open squares show the averaged {\sc Hi} data presented in
  \cite{widrow03}.  
  The lines are model fits for a three component model, fit to all the
  corrected PNe beyond 3.36\,kpc} 
\label{rot}
\end{figure}

The fully corrected rotation curve is shown in Fig.\ref{rot},
alongside the averaged {\sc Hi} rotation curve given in
\cite{widrow03}. With the exception of the first data point (where a
number of bulge PNe are likely to be present and hence the asymmetric
drift correction breaks down) it is clear that the PN.S data are in
very good agreement with the {\sc Hi} rotation curve and are of very
high quality out to $\sim$\,20\,kpc. Beyond this point there is
marginal evidence that the PN rotation curve drops to a lower velocity
than is seen in the {\sc Hi} data. However, there are very few PNe
per ring beyond 20\,kpc and we may simply be seeing halo contamination.

Figure \ref{rot} also shows a three component model fit to the PN
data from all but the first ring. This shows that the PN rotation
curve just reaches a radius where the dark matter halo becomes dominant.

\section{PNe in Halo Structures}

\begin{figure}[t!]
\begin{center}
\includegraphics[width=\textwidth]{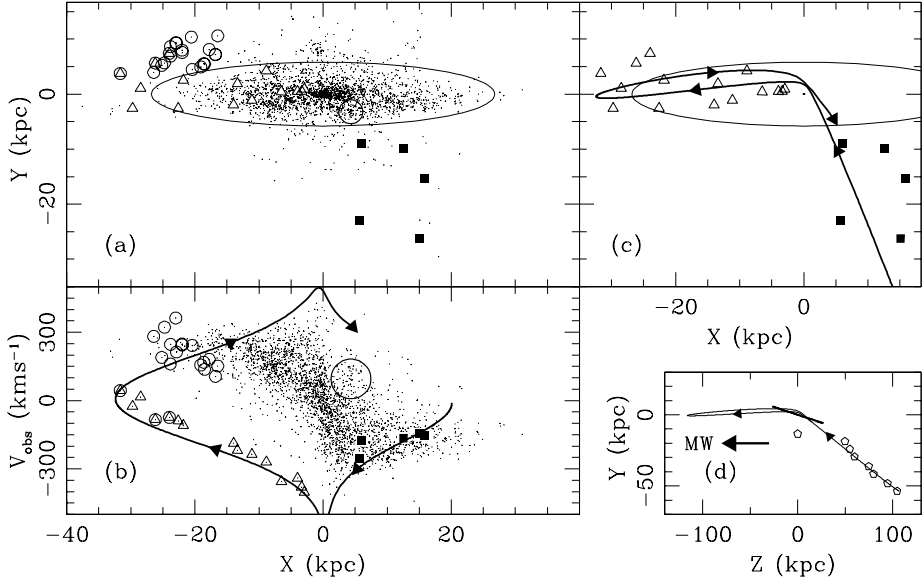}
\end{center}
\caption[]{Velocity Structures.
  \textbf{(a)} PNe positions. To the lower right (\emph{squares}) are
  PNe in the region of the Southern Stream, with velocities below
  -130\,kms$^{-1}$; to the upper left (\emph{circles}) are PNe in the
  Northern Spur region; and scattered through the left side of the
  disk (\emph{triangles}) are PNe selected for their unusual
  velocities. The ellipse marks the $2^\circ$ radius of the disk and
  the large circle marks the position of M32. 
  \textbf{(b)} Velocity versus distance along the major axis and the orbit
  location; PNe are marked as above.   
  \textbf{(c)} Orbit location in the $X-Y$ plane.
  \textbf{(d)} Orbit location in the $Z-Y$ plane. The arrow indicates
  the direction of the Milky Way; the short heavy line is the location
  of M31; and the points (\emph{pentagons}) show the measured depth to
  the stream from \cite{mcconnachie03}. 
} 
\label{stream}
\end{figure}

A number of PNe from this survey lie in the region of the Northern
Spur (\emph{circles}, Fig.\ref{stream}a), an over-density
of stars in Andromeda's halo \cite{ferg02}. This over-density is also 
seen in the PNe counts (similarly positioned halo fields in the
opposite quadrant have 1 or 2 PNe, while Northern Spur fields contain up to
$\sim$\,6). These planetaries have velocities similar to the maximum
observed rotation velocities in the disk (\emph{circles},
Fig.\ref{stream}b). This would suggest that the Northern Spur is not a
kinematically distinct substructure, but in fact forms part
of the disk, implying that the stellar disk must be severely warped.

PNe in the region of the Southern Stream \cite{ferg02} have also been
targeted by this survey (\emph{squares}, Figs.\ref{stream}a). Those PNe with
velocities below -130\,kms$^{-1}$ seem to form a coherent kinematic
structure (\emph{squares}, Fig.\ref{stream}b).

\section{Substructure in the Velocity Field}

Clearly a number of PNe in the sample have velocities which imply that
they are not just normal disk PNe.
One such population has been identified whereby the PNe lie below a line for
which there is no equivalent population in the opposite side of
the disk (\emph{triangles}, Fig.\ref{stream}b). 

These PNe have a low dispersion ($\sim$\,23\,kms$^{-1}$, including an
instrumental dispersion of $\sim$\,15-20\,kms$^{-1}$) about a straight
line fit. Such low dispersions are indicative of tidal streams.  
We therefore propose that these PNe are associated with a continuation of
the Southern Stream as it passes into the disk of M31, where it can no
longer be tracked via photometric methods (see also \cite{merrett03}).

The locations of the unusual PNe suggest a tidal stream lying almost
straight along the major axis. We have therefore generated a simple
orbit model (shown in Figs.\ref{stream}b, c and d) using a flattened
singular isothermal potential, $\Phi(R,z) = \frac{1}{2}v_c^2
\ln\left(R^2 + \frac{z^2}{q^2}\right)$ where $R$ and $z$ are polar
coordinates aligned with M31's disk plane; $v_c$ is set to the upper
envelope of PNe velocities (250\,kms$^{-1}$); and a value of $q=0.9$ is
adopted for the flattening. The orbit has been set to fall in along
the path defined in \cite{mcconnachie03} (Fig.\ref{stream}d), and turn
into the disk. This orbit agrees reasonably well with the velocities
of RGB stars in the Southern Stream region \cite{ibata04} and clearly
picks up the unusual PN velocities we have detected. 

The change in projection angle seen in Fig.\ref{stream}d between the
orbit's entry into the disk and as it moves across the disk, can explain the
significantly higher number of Stream PNe seen in the disk portion of
the orbit (15\,PNe) compared to the southern region (probably $\sim$\,5).

\section{Conclusion}

We have mapped the dynamics of the Andromeda galaxy's planetary nebula
system for PNe some 4.5\,mags into the luminosity function. From this
we have constructed a stellar rotation curve which agrees well with
the {\sc Hi} data. A simple three component model has been fit to the
PN data out to the realm where the dark halo becomes dominant. It is
important to measure this using the stellar population, as there can
be no ambiguity as to the nature of the force producing the
rotation---unlike gas, stars are subject only to gravitational
forces. 

Velocities have been measured for the halo structure known as the
Northern Spur, supporting the hypothesis that it represents a severe
warp in the stellar disk.

A number of PNe have been identified as forming a substructure within the
velocity field. These have a very low velocity dispersion, suggesting they are 
part of a tidal stream. We have shown they can be modeled as a
continuation of the Southern Stream as it turns into M31's disk where
it would no longer be detectable via photometric methods.

%


\begin{thebibliography}{8.}
\addcontentsline{toc}{section}{References}
\bibitem{bottema93} R. Bottema: A\&A \textbf{275}, 16 (1993)
\bibitem{braun91} R. Braun: Ap.J \textbf{372}, 54-66 (1991)
\bibitem{ferg02} A. Ferguson et al.: A.J. \textbf{124}, 1452-1463 (2002)
\bibitem{ibata04} R. Ibata et al.: astro-ph/0403068 (2004)
\bibitem{mcconnachie03} A. McConnachie et al.: MNRAS \textbf{343},
    1335-1340 (2003) 
\bibitem{merrett03} H. R. Merrett et al.: MNRAS \textbf{346}, L62-66 (2003)
\bibitem{neistein99} E. Neistein et al.: AJ \textbf{117}, 2666-2675 (1999)
\bibitem{perrett02} K. Perrett et al.: AJ, \textbf{123}, 2490-2510 (2002)
\bibitem{reitzel02} D. Reitzel and P. Guhathakurta: AJ, \textbf{124}, 234-265 (2002)
\bibitem{rubin70} V. Rubin and W. Ford: ApJ \textbf{159}, 379 (1970)
\bibitem{widrow03} L. Widrow et al.: ApJ \textbf{588}, 311-325 (2003)



\end{thebibliography}
\end{document}